\documentclass[epjc3]{svjour3}  

\usepackage{physics}
\usepackage{tikz}
\usetikzlibrary{decorations.markings,arrows,calc,patterns}
\usepackage{tikz-feynman}
\usepackage{pgfplots}
\usepackage{slashed}

\usepackage{hyperref}

\definecolor{darkBlue}{rgb}{0, 0, 0.8}

\hypersetup{
  bookmarksopen=true,     % open bookmarks
  colorlinks=true,        % false: boxed links; true: colored links
  linkcolor=darkBlue,     % color of internal links
  citecolor=darkBlue,      % color of links to bibliography
  filecolor=darkBlue,      % color of file links
  urlcolor=darkBlue        % color of external links
}

\renewcommand{\i}{\mathrm{i}}
\newcommand{\amp}{\mathcal{A}}
\newcommand{\SU}{\mathrm{SU}}
\newcommand{\ep}{\epsilon}
\newcommand{\vep}{\varepsilon}
\newcommand{\cd}{\cdot}
\newcommand{\Disc}{\text{Disc}}
\newcommand{\fsl}[1]{\slashed{#1}}

\def\centeron#1#2{{\setbox0=\hbox{#1}\setbox1=\hbox{#2}\ifdim
   \wd1>\wd0\kern.5\wd1\kern-.5\wd0\fi
   \copy0\kern-.5\wd0\kern-.5\wd1\copy1\ifdim\wd0>\wd1
   \kern.5\wd0\kern-.5\wd1\fi}}
\begin{document}

\title{Using Covariant Polarisation Sums in QCD}

\author{M. Kachelrie{\ss}   \and  M.N.~Malmquist}

\institute{Institutt for fysikk, NTNU, Trondheim, Norway}

\date{Received: date / Accepted: date}

\maketitle

\begin{abstract}
Covariant gauges lead to spurious, non-physical polarisation states of
gauge bosons. In QED,  the use of the Feynman gauge,
$\sum_{\lambda} \vep_\mu^{(\lambda)}\vep_\nu^{(\lambda)\ast} = -\eta_{\mu\nu}$,
is justified by the Ward identity which ensures that the contributions of
non-physical polarisation states cancel in physical observables. In contrast,
the same replacement
can be applied only to a single external gauge boson in squared amplitudes of
non-abelian gauge theories like QCD.
In general, the use of this replacement requires to include
external Faddeev-Popov ghosts. We present a pedagogical derivation of  these
ghost contributions  applying the optical theorem and the Cutkosky cutting
rules. We find that the resulting cross terms
$\amp(c_1,\bar{c}_1;\ldots)\amp(\bar{c}_1,c_1;\ldots)^\ast$
between ghost amplitudes cannot be transformed into
$(-1)^{n/2}|\amp(c_1,\bar{c}_1;\ldots)|^2$ in the case of more
than two ghosts.
Thus the Feynman rule stated in the literature holds only for
two external ghosts, while it is in general incorrect.
\end{abstract}

\section{Introduction}
\label{sec:intro}

The traditional way to derive physical observables like decay widths
and cross sections from Feynman amplitudes $\amp$ is to calculate the
squared amplitude $|\amp|^2$ and then to sum over the final and to average
over the initial spin and polarisation states of the external particles.
In the case of fermions, one obtains for each sum a projection operator
on the positive or negative energy solutions of the Dirac
equations, and a fermion line is thereby converted into a
trace of spinors which is relatively easy to calculate.
For particles with spin $s\geq 1$,
the summation over polarisation states is complicated  by the fact there is
a mismatch between the physical number of polarisation states and 
field variables. In particular, a massless spin-one field $A^\mu$
has four components but only two physical  polarisation states.
Excluding the non-physical states requires to choose a non-covariant
form of the polarisation sum, what in turn results in lengthy
intermediate expressions for the squared amplitude $|\amp|^2$. For instance,
the squared amplitude of the four-gluon amplitude in QCD contains
228\,420 terms before simplifications reduce them to four.

In the case of photons, one can avoid this complication by choosing
for the sum over polarisation states a covariant gauge as, e.g., the
Feynman gauge, $\sum_{\lambda} \vep_\mu^{(\lambda)}\vep_\nu^{(\lambda)\ast} = -\eta_{\mu\nu}$. Although  both
time-like and longitudinal photons are included in this polarisation sum,
the Feynman gauge leads in QED to the same results as a physical gauge
where only transversely polarised photons propagate: The photon
couples to a conserved current, and thus the resulting Ward identity,
$k_\mu \amp^\mu=0$ for an external photon with four-momentum $k^\mu$, ensures
that the contributions of the two redundant
polarisation states cancel in physical observables.

In the non-abelian case, the gauge boson couples to a current
which is the sum of the conserved fermion current and a piece
induced by the self-coupling of the gauge field. This current
is neither gauge-invariant nor conserved. Therefore the simple
QED Ward identity does not hold. In particular, the
contribution of the two redundant polarisation states
do not cancel any more, if covariant polarisation sums are used
for more than one external gauge boson. If one insists to use the
simpler covariant gauges, one has therefore to add Faddeev-Popov
ghosts.
In the case of internal gluons, this approach is the standard
procedure applied in almost all loop calculations.  The use of covariant
gauges and  Faddeev-Popov ghosts for
external  gauge bosons is, on the contrary, rarely employed.
An exception is the work of  Cutler and Sivers~\cite{Cutler:1977qm}
who performed the first calculation of four-gluon scattering in QCD.
Their result differs however from the now accepted one, first found in the
same year by Combridge, Kripfganz and Ranft using a different
method~\cite{Combridge:1977dm}. Employing  \(- \eta^{\mu \nu}\) as
polarisation sum method in QCD is also described briefly in Nachtmann's
textbook~\cite{Nachtmann:1990ta}. There it is stated that the
covariant polarisation sum can be used, given that we also add squared
ghost amplitudes, modified by factors of \({(-1)}^{n}\) for \(2n\)
external Faddeev-Popov ghosts. All possible amplitudes replacing some
even number of external gluons by ghosts should be included.

In this work, we examine  the use of covariant gauges for external
gauge bosons and in particular the Feynman rules for external ghosts.
The possibility to use covariant gauges also for the polarisation sums
of external gluons adding also amplitudes with external ghosts
is guaranteed by the optical theorem or, equivalently, by  the
invariance of the amplitude under gauge transformations of internal
lines. Applying the cutting rules of Cutkosky and Landau, we can connect
in turn the properties of internal lines to those of  external states.
We first introduce these cutting rules in a QED example in
Section~\ref{sec:conn-optic-theor}. This example contains most of the
concepts needed to understand also the QCD case. The additional
subtleties arising in QCD are treated in Section~\ref{sec:qcd-case}.
Finally, we apply the method to four-gluon scattering in
Section~\ref{sec:cutting-more-than}. This example demonstrates that, when
more than two polarisation sums are replaced by \(-\eta^{\mu \nu}\),
the ghost contributions can no longer be expressed as squared amplitudes
times a sign factor. Instead the cross terms between ghost
amplitudes and their \(c \leftrightarrow \bar{c}\) counterpart
have to  be used. This implies that the generalisation
of the Feynman rule for external ghosts derived from the $\bar qq\to gg$
amplitude using squared ghost amplitudes is incorrect.
Finally, we summarize in Section~\ref{summary}.

\section{Cutting Rules in a QED Example}
\label{sec:conn-optic-theor}

A well-known consequence of the unitarity of the S-matrix is the
optical theorem~\cite{Peskin:1995ev,Kachelriess:2017cfe},
\begin{equation}
  \label{eq:27-ymsa}
  2 \Im{\amp(i \to i)} = \sum_{N = 2}^{\infty} 
  \sum_{\xi_1 \dotsc \xi_N} \int |\amp(i \to \xi_1 \dotsc \xi_N)|^2 \dd{\Phi^{(N)}}.
\end{equation}
It relates the imaginary part of the forward scattering amplitude
$\amp(i \to i)$ to
the total cross section of the initial state scattering into any final
state. The sum over $N$ has to be truncated at a finite \(N\) corresponding
to the considered order in perturbation theory, while the sum over
\(\xi_i\) includes all quantum numbers describing the final state. 

\begin{figure}[ht]
  \centering
  \includegraphics{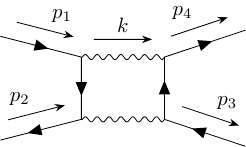}
  \caption{The diagram \(\amp^{(\gamma \gamma)}\) contributing to the
    process \(e^- e^+ \to e^- e^+\) at the one-loop level.}
  \label{fig:qed-example}
\end{figure}
We consider the optical theorem in the specific example of \(e^- e^+
\to e^- e^+\) scattering at order \(\alpha^2\). The imaginary part of the
amplitude for this process is the sum of the imaginary parts of the
contributing Feynman diagrams. We will consider only the one-loop diagram 
shown in Figure~\ref{fig:qed-example} in detail. For later convenience, we
choose to represent the diagram in terms of two off-shell
sub-amplitudes as follows,
\begin{align}
  \label{eq:amp-qed-example-simplified-form}
  \i \amp^{(\gamma \gamma)}
  =
  \int \frac{\dd[4]{k}}{(2\pi)^4} &
  \i\amp_{\mu \nu}^{(u)}(e^-(p_1) e^+(p_2) \to \gamma(p_1 + p_2 -
  k)\gamma(k))
  \\
  \nonumber
  \cross &
  \i\amp_{\nu' \mu'}^{(u)}(\gamma(k)\gamma(p_1 + p_2 - k) \to e^+(p_3) e^-(p_4))
  \\
  \nonumber
  \cross &
  \frac{\i N^{\nu \nu'}(k)}{k^2 + \i \ep} 
  \cross \frac{\i N^{\mu \mu'}(p_1 + p_2 - k)}{{(p_1 + p_2 - k)}^2 +
    \i \ep}.
\end{align}
Here, \(\i\amp^{(\gamma \gamma)}\) corresponds to the diagram in
Figure~\ref{fig:qed-example}, while \(\i\amp_{\mu \nu}^{(u)}\) represents
the \(u\)-channel diagram for either \(e^- e^+ \to \gamma \gamma\)
or \(\gamma \gamma \to e^- e^+\) scattering with the external polarisation
vectors removed. The photon momenta \(k\) and \(p_1 + p_2 - k\) in the
latter two amplitudes are off-shell. Finally, \(N^{\mu \nu}\) denotes
the  numerator of a photon propagator. After seeing how the imaginary part
can be computed for this diagram, it should be clear how the same
method can be applied to the remaining one-loop diagrams for \(e^- e^+
\to e^- e^+\).

The imaginary part of a Feynman diagram is proportional to the
discontinuity across its \(s\)-channel branch
cut~\cite{Peskin:1995ev,Kachelriess:2017cfe}.
Such discontinuities can be computed elegantly
by using the cutting rules of Cutkosky and
Landau~\cite{Cutkosky:1960sp,Landau:1959fi}. We illustrate how the
rules work in our present example; the diagram of
Figure~\ref{fig:qed-example}. First we should find all sets of
internal lines such that the corresponding momenta \(q_i\) satisfy \(s
= (\sum_i q_i)^2\). Additionally, every pair of lines \(\{q_i, q_j\}\) must
share at least one common closed loop. In our example we see that the
set of internal lines with momenta \(k\) and \(p_1 + p_2 - k\) is the
only set that works. Next, the Cutkosky rules tell us that to compute the
discontinuity we should perform the replacement
\begin{equation}
  \label{eq:cutkosky-rule}
  \frac{1}{q_i^2 - M_i^2} \to 2 \pi \i \delta^{(+)}(q_i^2 - M_i^2),
\end{equation}
in each of the propagators corresponding to our chosen internal
lines. Here \(\delta^{(+)}\) means we take only the principal root of
\(q_i^2 = M_i^2\). From our expression
(\ref{eq:amp-qed-example-simplified-form}), now specialized to forward
scattering, we obtain then
\begin{align}
  \label{eq:imaginary-part-of-qed-example}
  2 \i \Im \big\{ \amp^{(\gamma \gamma)} \big\} &=
  \Disc_{s\text{-channel}}\big\{ \amp^{(\gamma \gamma)} \big\}
  \\
  \nonumber
  &= \i \int \frac{\dd[4]{k}}{(2\pi)^2}
  \delta^{(+)}(k^2) \delta^{(+)}({(p_1 + p_2 - k)}^2)
  \\
  \nonumber
  & \hphantom{=} \cross
  \amp_{\mu \nu}^{(u)}(e^-(p_1) e^+(p_2) \to \gamma(p_1 + p_2 - k)\gamma(k))
  \\
  \nonumber
  & \hphantom{=} \cross
  \amp_{\nu' \mu'}^{(u)}(\gamma(k)\gamma(p_1 + p_2 - k) \to e^+(p_2) e^-(p_1))
  \\
  \nonumber
  & \hphantom{=} \cross
  N^{\nu \nu'}(k)
  N^{\mu \mu'}(p_1 + p_2 - k).
\end{align}

The way we have chosen to express \(\amp^{(\gamma \gamma)}\) makes it
clear that its form is already quite close to the one of the optical
theorem. The
\(\delta^{(+)}(k^2)\) sets the external photon momentum \(k\)
on-shell in the two sub-amplitudes. The second \(\delta^{(+)}({(p_1 +
  p_2 - k)}^2)\) sets the other external photon on-shell. The loop
integral is transformed by the delta functions into the two-particle
phase-space integral. To be precise,
\begin{equation*}
  \int \frac{\dd[4]{k}}{(2\pi)^2} \delta^{(+)}(k^2) 
  \delta^{(+)}({(p_1 + p_2 - k)}^2)
  =
  \int \frac{\dd[3]{k}}{(2 \pi)^3 2 \omega_k} 
  \frac{\dd[3]{k'}}{(2 \pi)^3 2 \omega_{k'}} {(2 \pi)}^4
  \delta^{(4)}(p_1 + p_2 - k - k'),
\end{equation*}
where the right-hand side is twice the phase-space integral of two
photons, since they are identical particles. This factor of two means
that the single loop diagram in Figure~\ref{fig:qed-example} gives
rise to the squares of both the \(u\)- and \(t\)-channel tree diagrams
for \(e^+ e^- \to \gamma \gamma\). This holds since we can turn the
\(u\)-channel sub-amplitudes above into \(t\)-channel sub-amplitudes
by interchanging \(k\) and \(k'\), which does not change the value of
the integral.

Although close, we have not quite arrived at the squared amplitude
expected from the optical theorem yet. Firstly, we need to realize that
\begin{equation}
\label{eq:complex-ocnjugating-qed-amp}
\amp_{\nu' \mu'}^{(u)}(\gamma(k)\gamma(k') \to e^+(p_2) e^-(p_1))
=
\amp_{\mu' \nu'}^{(u)}(e^-(p_1) e^+(p_2) \to \gamma(k')\gamma(k))^*.
\end{equation}
This follows from the property that complex conjugation of the
amplitude turns an incoming physical particle into an outgoing one and vice
versa. Technically, this is facilitated for a Dirac chain by relations like
\((\bar{u} \gamma^{\mu_1} \dotsc \gamma^{\mu_n} v)^* = \bar{v}
\gamma^{\mu_n} \dotsc \gamma^{\mu_1} u\). Using this relation, we obtain
\begin{align}
  \label{eq:imaginary-part-of-qed-example-final-form}
  2 \Im \big\{ \amp^{(\gamma \gamma)} \big\} =
  2 \int \dd{\Phi^{(2)}} &
  \amp_{\mu \nu}^{(u)}(e^-(p_1) e^+(p_2) \to \gamma(k')\gamma(k))
  \\
  \nonumber
  \cross &
  N^{\nu \nu'}(k)
  N^{\mu \mu'}(k')
  \amp_{\mu' \nu'}^{(u)}(e^-(p_1) e^+(p_2) \to \gamma(k')\gamma(k))^*.
\end{align}
The right hand side of this equation is the sum of the squared \(u\)-
and \(t\)-channel diagrams for \(e^+ e^- \to \gamma \gamma\),
integrated over the two particle phase space and with the sum over
polarisations replaced by \(N^{\mu \mu'}(k)\). Had we repeated the
analysis for the one-loop diagram similar to
Figure~\ref{fig:qed-example} but with the photon lines crossed, we
would have found the cross term between the \(u\)- and \(t\)-channel
diagrams, again with the sum over polarisations replaced by
\(N^{\mu\mu'}(k)\). Thus the optical theorem~(\ref{eq:27-ymsa})
connects the numerator of the photon propagator with the sum over
polarisations of external photons. In particular, the optical theorem
implies that the Feynman propagator \(N^{\mu \mu'}(k) = - \eta^{\mu
  \mu'}\) can be used to sum over polarisations.

The sum over transverse, physical polarisations neither equals  \(-
\eta^{\mu \mu'}\) nor it is gauge dependent like the numerator of
the propagator  \(N^{\mu \mu'}(k)\). The optical theorem then requires
the amplitudes
to have such a form that any non-transverse part of \(N^{\mu
  \mu'}(k)\) is cancelled. This is equivalent to the freedom that
we may use the propagator in any gauge in internal lines. The connection
between the covariant and physical polarisation sum is
\begin{equation}
  \label{eq:polsum-and-propagator-relation}
  - \eta^{\mu \nu} = 
  \sum_\lambda \vep_\lambda^{\mu *} \vep_\lambda^{\nu}
  - \frac{k^\mu \tilde{k}^\nu + \tilde{k}^\mu k^\nu}{2 \omega_{k}},
\end{equation}
where \(\tilde{k}^\mu = (\omega_k, - \vb{k})\) and \(\lambda\) are the
two transverse polarisation states~\cite{Kachelriess:2017cfe}. Thus the Ward
identity \(k^\mu \amp_\mu = 0\) is the required property,
ensuring that only the transverse part of \(N^{\mu \mu'}(k)\)
contributes.

\section{Extension to QCD}
\label{sec:qcd-case}

\begin{figure}[ht]
  \centering
  \includegraphics{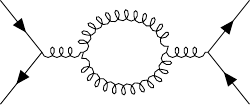}
  \caption{The diagram \(\amp^{(g\text{-loop})}\) contributing to
    \(\bar{q} q \to \bar{q} q\) at the one-loop level.}
  \label{fig:qcd-example}
\end{figure}

We now turn to the equivalent QCD process \(\bar{q} q \to \bar{q} q\)
in order to see how the discussion of the previous section has to be
changed in the case of a non-abelian gauge theory.
The QED-like diagrams of the previous section contribute now as before,
being only altered by the added \(\SU(3)\) colour
structure. Since the kinematics is the same, twice the imaginary part
of the QCD equivalent of Figure~\ref{fig:qed-example} again gives the
squares of the \(t\)- and \(u\)-channel diagrams, now for \(\bar{q} q
\to gg\). However, the \(\bar{q} q \to gg\) process has an additional diagram
compared to its QED counterpart: the \(s\)-channel diagram. In
correspondence with the optical theorem we have the diagram in
Figure~\ref{fig:qcd-example}, whose imaginary part gives the square of
that \(s\)-channel diagram. Additionally, we have two one-loop diagrams
with a single three-gluon vertex that give the cross terms between the
\(s\)-channel and the \(t\)- and \(u\)-channel diagrams.

Repeating the argument of the previous section to find the imaginary
part of the diagram in Figure~\ref{fig:qcd-example} we are left with a
difficulty. Again we should be able to use any gauge for the internal
gluon propagator numerators \(\i \delta^{ab} N^{\mu \mu'}(k)\), but
now the QED Ward identity does not hold. The non-transverse part of \(N^{\mu
  \mu'}(k)\) gives a non-zero contribution when contracted into the
two sub-amplitudes of the cut. This non-transverse contribution is not
present on the right-hand side of the optical theorem
(\ref{eq:27-ymsa}), so we have an apparent contradiction.

\begin{figure}[ht]
  \centering
  \includegraphics{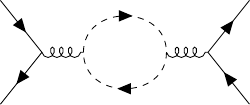}
  \caption{The diagram with Faddeev-Popov ghosts contributing to
    \(\bar{q} q \to \bar{q} q\) at the one-loop level.}
  \label{fig:qcd-example-ghost-loop}
\end{figure}

We consider this non-transverse contribution in detail in the Feynman
gauge, following the same steps as in the QED example. Inserting the
expression~(\ref{eq:polsum-and-propagator-relation}) for \(N^{\mu \mu'} = -
\eta^{\mu \mu'}\) we obtain for the imaginary part of the diagram in
Figure~\ref{fig:qcd-example},
\begin{align}
  \label{eq:imaginary-part-qcd-example}
  2 \Im{\amp^{(g\text{-loop})}} =&
  \sum_{a, a'} \sum_{\lambda \lambda'} \int \dd{\Phi^{(2)}}
  \abs{ \amp^{(s\text{-channel})}(\bar{q}q \to gg) }^2
  \\
  \nonumber
  & + 2 \int \dd{\Phi^{(2)}}
  \left[
    g_{\rm s}^2 \frac{\i f^{baa'} T^b_{ij}}{(p_1 + p_2)^2}
    \bar{v}(2)\fsl{k} u(1)
  \right]
  \left[
    g_{\rm s}^2 \frac{\i f^{baa'} T^b_{ji}}{(p_1 + p_2)^2}
    \bar{v}(2) (- \fsl{k}') u(1)
  \right].
\end{align}
Here, $g_{\rm s}$ is the strong coupling, \(\lambda, \lambda'\) are the
physical polarisation states of the
two final state gluons, \(a, a'\) are their colour states and \(k, k'\)
are their momenta. Now consider the following amplitudes involving
external Faddeev-Popov ghosts, which we denote by \(c\) and
\(\bar{c}\),
\begin{align}
  \label{eq:ghost-amplitudes-in-cancellation-1}
  \amp(q(p_1) \bar{q}(p_2) \to c(k) \bar{c}(k')) &= g_{\rm s}^2 \frac{\i f^{baa'} T^b_{ij}}{(p_1 + p_2)^2}
    \bar{v}(2)\fsl{k} u(1), \\
  \label{eq:ghost-amplitudes-in-cancellation-2}
  \amp(c(k) \bar{c}(k') \to q(p_1) \bar{q}(p_2)) &= g_{\rm s}^2 \frac{\i f^{baa'} T^b_{ji}}{(p_1 + p_2)^2}
    \bar{v}(2) (- \fsl{k}') u(1).
\end{align}
Remembering also that ghosts are fermions and thus the ghost loop contains
an additional factor of
\(-1\), we see that the non-transverse contribution on the second line
of (\ref{eq:imaginary-part-qcd-example}) is exactly minus the cut of
the diagram in Figure~\ref{fig:qcd-example-ghost-loop}. The factor of
two accounts for the fact that, unlike the final state gluons, \(c\)
and \(\bar{c}\) are not identical.

As expected, the optical theorem is saved by the inclusion of Faddeev-Popov
ghosts. Thus covariant propagator numerators can, like in the previous section,
be used to sum over polarisation, if we also include
ghosts. For the present case of \(q(p_1) \bar{q}(p_2) \to g(k) g(k')\)
the relevant ghost term is
\begin{align}
  \label{eq:ghost-term-qcd-example}
  & - 2 \amp(q(p_1) \bar{q}(p_2) \to c(k) \bar{c}(k')) 
  \amp(c(k) \bar{c}(k') \to q(p_1) \bar{q}(p_2)) \\
  & =
  \nonumber
  2 \amp(q(p_1) \bar{q}(p_2) \to c(k) \bar{c}(k')) 
  \amp(q(p_1) \bar{q}(p_2) \to \bar{c}(k) c(k'))^*,
\end{align}
which follows from the combination of
(\ref{eq:imaginary-part-qcd-example}) with
(\ref{eq:ghost-amplitudes-in-cancellation-1}) and
(\ref{eq:ghost-amplitudes-in-cancellation-2}). Note that unlike before
we no longer get a square, but rather the cross term between a ghost
amplitude and its \(c \leftrightarrow \bar{c}\) counterpart. For
physical particles complex conjugation exchanges incoming and outgoing
particles, like we saw for Dirac fermions in the previous
section. While complex conjugating equation
(\ref{eq:ghost-amplitudes-in-cancellation-2}) gives the correct colour
structure for exchanging incoming and outgoing particles due to
\(T_{ij}^{a*} = T^a_{ji}\), the \(- \fsl{k}'\) from the ghost vertex
is unchanged. Reinterpreting the ghosts as outgoing we then get an
additional sign and the final state is \(\bar{c}(k) c(k')\), not the
\(c(k) \bar{c}(k')\) that would give a square.

This example with only two cut gluons allows for a peculiar,
further transformation of the result in Eq.~(\ref{eq:ghost-term-qcd-example}),
since the ghost momenta $\fsl{k}$ sits next to an on-shell spinor:
Using  the Dirac equation to transform
\(\bar{v}(2) (- \fsl{k}') u(1) = \bar{v}(2) \fsl{k} u(1)\) in
(\ref{eq:ghost-amplitudes-in-cancellation-2}), we can write the ghost
term (\ref{eq:ghost-term-qcd-example}) as a square with a sign \(-1\)
relative to our term. Such a transformation is clearly not possible in
general, as we will also see in the next section in a concrete example.

\section{Cutting More Than Two Gluons}
\label{sec:cutting-more-than}

Now that we understand how a covariant propagator numerator can be used
to sum over polarisations, we will apply the method to a final example. We
will illustrate that there is no need to first consider the loop amplitudes
in order to write down the relevant ghost terms. By choosing the
four-gluon amplitude we will also be able to see how the discussion of
the previous section generalizes to more than two cut gluons\footnote{A
  FORM file for
computing the three terms required for using the Feynman gauge on all
external lines in the four gluon amplitude can be found at
\url{https://gitlab.com/magnunm/yang-mills-scattering-amplitudes/-/blob/master/gluon-gluon-scattering/4-feynman-gauge.frm}. We
have also written a Mathematica program that illustrates in a
pedagogical manner the ideas presented in the text with concrete
calculations in the four gluon case, it can be found at
\url{https://gitlab.com/magnunm/yang-mills-scattering-amplitudes/-/blob/master/gluon-gluon-scattering/gluon-gluon-scattering-ghost-cancellation.wl}}.
To gradually increase the complexity we will start by using \(-
\eta^{\mu \nu}\) as the polarisation sum for three of the gluons,
before moving on to all four. We adopt a notation where \(g\) stands
for a gluon which we sum over purely transverse polarisations, while
\(\tilde{g}\) is a gluon where we replace the sum over polarisations
by \(- \eta^{\mu \nu}\). Keeping one gluon transverse we then find
\begin{align}
  \label{eq:gg-to-gg-three-long-gluon}
  \sum_{\text{pol.}} \tr \abs{\amp(g \tilde{g} \to \tilde{g} \tilde{g})}^2
  =& \sum_{\text{pol.}} \tr \abs{\amp(g g \to g g)}^2 \\
  \nonumber
  &+ 324 g_{\rm s}^4 - 288 g_{\rm s}^4 \frac{st}{u^2} - 288 g_{\rm s}^4 \frac{su}{t^2} - 432 g_{\rm s}^4 \frac{tu}{s^2},
\end{align}
where $s,t$ and $u$ denote the usual Mandelstam variables.
There are now three possible positions for a ghost \(c\), and having
fixed that there are two remaining positions for the second external
ghost. In other words there are \(3 \cd 2 = 6\) possible ghost
amplitudes. Of the six corresponding ghost cross terms only three are
unique since the remaining three are related by a complex
conjugation. These three ghost cross terms are
\begin{align}
  \label{eq:gg-to-gg-three-long-ghost1}
  \sum_{\text{pol.}} \tr \amp(g \tilde{g} \to \bar{c} c)
  \amp(g \tilde{g} \to c \bar{c})^*
  &= - 90 g_{\rm s}^4 + 216 g_{\rm s}^4 \frac{tu}{s^2}, \\
  \label{eq:gg-to-gg-three-long-ghost2}
  \sum_{\text{pol.}} \tr \amp(g c \to \tilde{g} c)
  \amp(g \bar{c} \to \tilde{g} \bar{c})^*
  &= - 18 g_{\rm s}^4 + 36 g_{\rm s}^4 \frac{t}{s} + 144 g_{\rm s}^4 \frac{su}{t^2}, \\
  \label{eq:gg-to-gg-three-long-ghost3}
  \sum_{\text{pol.}} \tr \amp(g c \to c \tilde{g})
  \amp(g \bar{c} \to \bar{c} \tilde{g})^*
  &= - 54 g_{\rm s}^4 - 36 g_{\rm s}^4 \frac{t}{s} + 144 g_{\rm s}^4 \frac{st}{u^2}.
\end{align}
Summing over all the six ghost cross terms gives minus the second line
of (\ref{eq:gg-to-gg-three-long-gluon}), as expected from the optical
theorem.

Computing next the square
\begin{equation}
  \sum_{\text{pol.}} \tr \abs{\amp(g \tilde{g} \to \bar{c} c)}^2
  = 108 g_{\rm s}^4 + 36 g_{\rm s}^4 \frac{t}{s} - 216 g_{\rm s}^4 \frac{tu}{s^2},
\end{equation}
and comparing it to (\ref{eq:gg-to-gg-three-long-ghost1}), we see that
the two expressions no longer have a simple relationship: Writing the ghost
terms arising in the optical theorem as squares is not possible when one
goes beyond two cut gluons.

Finally, we substitute \(-\eta^{\mu \nu}\) for the sum over
polarisations for all four external gluons. We must then also include
all amplitudes with four external ghosts. There are six such
amplitudes and six corresponding cross terms. As before only three
need to be computed, the remaining three being complex conjugates of
the first ones. Now there are twelve amplitudes with two external ghosts.
The six amplitudes in
(\ref{eq:gg-to-gg-three-long-ghost1})--(\ref{eq:gg-to-gg-three-long-ghost3})
plus an additional six where there is a ghost at position \(1\). The
last six are however just given by relabellings of momenta in the
first. The full equality implied by unitarity is then
\begin{align}
  \label{eq:4-gluon-equality-unitatiry}
  \sum_{\text{pol.}} \tr \abs{\amp(g g \to g
    g)}^2
  &=
  \sum_{\text{pol.}} \tr \abs{\amp(\tilde{g} \tilde{g} \to \tilde{g}
    \tilde{g})}^2 \\
  \nonumber
  + 4
  \sum_{\text{pol.}} \tr \amp(\tilde{g} \tilde{g} \to c \bar{c})
  \amp(\tilde{g} \tilde{g} \to \bar{c} c)^*
  &+ 4
  \sum_{\text{pol.}} \tr \amp(\tilde{g} c \to \tilde{g} c)
  \amp(\tilde{g} \bar{c} \to \tilde{g} \bar{c})^* \\
  \nonumber
  + 4
  \sum_{\text{pol.}} \tr \amp(\tilde{g} c \to c \tilde{g})
  \amp(\tilde{g} \bar{c} \to \bar{c} \tilde{g})^*
  &+ 2
  \sum_{\text{pol.}} \tr \amp(c \bar{c} \to \bar{c} c)
  \amp(\bar{c} c \to c \bar{c})^* \\
  \nonumber
  + 2
  \sum_{\text{pol.}} \tr \amp(c \bar{c} \to c \bar{c})
  \amp(\bar{c} c \to \bar{c} c)^*
  &+ 2
  \sum_{\text{pol.}} \tr \amp(c c \to c c)
  \amp(\bar{c} \bar{c} \to \bar{c} \bar{c})^*.
\end{align}
Considering that crossing symmetry holds for these amplitudes, we can
get all the ghost terms above by crossing the expressions for
\(\amp(\tilde{g} \tilde{g} \to c \bar{c}) \amp(\tilde{g} \tilde{g} \to
\bar{c} c)^*\) and \(\amp(c \bar{c} \to \bar{c} c) \amp(\bar{c} c \to
c \bar{c})^*\) into the appropriate channels. The
amplitudes with four external ghosts arise from loop amplitudes with
two ghost loops. Since they also involve four ghosts being
reinterpreted as outgoing/incoming, the signs cancel out as with two
external ghosts. This clearly continues to any number of external
ghosts.

\section{Summary}
\label{summary}

We have reviewed the use of covariant gauges for the polarisation states of
external gauge bosons. While in QED the application of such gauges is
justified by the Ward identity, $k_\mu \amp^\mu=0$, the use of these gauges in
non-abelian gauge theories like QCD requires to include external
Faddeev-Popov ghosts. Since covariant gauges and  Faddeev-Popov
ghosts for external  gauge boson have been rarely employed in QCD,
it seems that the Feynman rules for external ghosts have not been previously
checked beyond the simplest case of two external ghosts.

We have used the optical theorem  and the cutting rules of Cutkosky
and Landau to show how the non-physical contributions due to
non-transverse polarisation states present in covariant gauges are
cancelled by the Faddeev-Popov ghosts. This cancellation requires to
take into account all possible amplitudes replacing an even number of
external gluons by ghosts. These amplitudes should be multiplied by the
complex conjugate of their \(c \leftrightarrow \bar{c}\) counterparts;
they have to be summed with no additional sign. Only in the  simplest
case of two external ghost is the transformation
$\amp(c,\bar{c})\amp(\bar{c},c_1)^\ast= -|\amp(c,\bar{c})|^2$ possible.
In the past, this peculiar case has been probably the basis to argue
that the general Feynman rule for external ghosts is
$(-1)^{n/2}|\amp(c_1,\bar{c}_1;\ldots)|^2$, as it is given, e.g., in the
textbook by  Nachtmann~\cite{Nachtmann:1990ta}. Using this wrong rule one
reproduces also the incorrect result
for $gg\to gg$ scattering obtained in Ref.~\cite{Cutler:1977qm}.

In this work, we have used the unitarity of the S-matrix as our starting
point. This allows one to derive the form of the ghost terms at any order of
perturbation theory. The existence of Slavnov-Taylor identities that
underlie the cancellation of unphysical degrees of freedom is in
this approach a consequence of the assumption of
unitarity. Alternatively, we could have started by deriving these
identities. Then the steps used here to find the ghost terms may be
generalised into 't Hooft's method of proving the unitarity of the
S-matrix~\cite{tHooft:1971akt}. Mathematically more sophisticated are proofs
based on the BRST-formalism as, e.g.,  the one presented by Kugo and
Ojima~\cite{Kugo:1979gm}. The more elementary diagrammatic approach
followed here has the advantage with respect to Ref.~\cite{Kugo:1979gm}
to expose explicitly the Feynman rules
required for a cancellation of external ghost particles.

Finally, we should warn the reader that there is a good reason why the
use of covariant gauges combined with ghosts for external gluons has
been unpopular: Firstly, the increase in the number of diagrams is not
really compensated by the simplicity of the polarisation sum in the Feynman
gauge. More importantly, the advent of new powerful methods to calculate
helicity amplitudes involving gauge bosons in the last two decades has
made this approach obsolete for most practical purposes.

\subsection*{Acknowledgments}

We would like to thank Jonas Tjemsland for useful comments.

%\nolinenumbers

%\bibliographystyle{spphys}
%\bibliography{literature.bib}

\end{document}